\documentclass[12pt]{article}

\usepackage{newtxtext,newtxmath}
\usepackage{longtable}
\usepackage{makecell}
\usepackage{multirow}
\usepackage{tabularx}
\usepackage{adjustbox}

\usepackage{booktabs}
\usepackage{graphicx}

\usepackage[letterpaper,margin=1in]{geometry}

\renewenvironment{abstract}
	{\quotation}
	{\endquotation}

\date{}

\makeatletter
\renewcommand{\fnum@figure}{\textbf{Figure \thefigure}}
\renewcommand{\fnum@table}{\textbf{Table \thetable}}
\makeatother

\usepackage{scicite}

\usepackage{url}

\def\scititle{
	Finer resolutions and targeted process representations in earth systems models improve hydrologic projections and hydroclimate impacts.
}
\title{\bfseries \boldmath \scititle }

\author
{Puja Das,$^{1\ast}$ Auroop R. Ganguly$^{1,2,3\ast}$\\
\\
\normalsize{$^{1}$Sustainability and Data Sciences Laboratory, Northeastern University, Boston, MA, USA.}\\
\normalsize{$^{2}$The Institute for Experiential AI and Roux Institute, Northeastern University, Boston, MA, USA.}\\
\normalsize{$^{3}$Pacific Northwest National Laboratory, Richland, WA, USA}\\
\\
\normalsize{$^\ast$To whom correspondence should be addressed; E-mail:  a.ganguly@northeastern.edu.}
}

\begin{document} 

% Insert the title and author list
\maketitle

% Abstract, in bold
% There are strict length limits, and not all formats have abstracts.
% Consult the journal instructions to authors for details.
% Do not cite any references in the abstract.
\begin{abstract} \bfseries \boldmath
% Start with one or two sentences of background
Earth system models inform water policy and interventions, but knowledge gaps in hydrologic representations limit the credibility of projections and impacts assessments. The literature does not provide conclusive evidence that incorporating higher resolutions, comprehensive process models, and latest parameterization schemes, will result in improvements. We compare hydroclimate representations and runoff projections across two generations of Coupled Modeling Intercomparison Project (CMIP) models, specifically, CMIP5 and CMIP6, with gridded runoff from Global Runoff Reconstruction (GRUN) and ECMWF Reanalysis V5 (ERA5) as benchmarks. Our results show that systematic embedding of the best available process models and parameterizations, together with finer resolutions, improve runoff projections with uncertainty characterizations in 30 of the largest rivers worldwide in a mechanistically explainable manner. The more skillful CMIP6 models suggest that, following the mid-range SSP370 emissions scenario, 40\% of the rivers will exhibit decreased runoff by 2100, impacting 260 million people.
\end{abstract}

% The first paragraph of any Science paper does NOT have a heading
% Nor is it indented
\noindent

Earth System Models (ESMs) are critical tools for understanding climate science, supporting climate adaptation, and informing water resources management. The latest generation of these ESMs enables more refined analysis of hydroclimate responses \cite{clark2015improving,lurton2020implementation,liu2023performance}, yet significant challenges remain in ensuring the credibility of their projections, particularly for hydrologic cycle components like runoff at stakeholder-relevant spatiotemporal resolutions \cite{hulme2009keeping,kumar2014regional}. As water is essential for human sustenance, agriculture, energy production, and ecosystem maintenance, accurate projections of the global hydrologic cycle (GHC) are crucial \cite{douville2021water}, especially given the centrality of water to all 17 United Nations sustainable development goals (UN SDGs)\cite{mugagga2016centrality}. However, integrating hydrologic and hydroclimate processes into ESMs get complicated by the heterogeneity of the process, data and  environment \cite{ganguly2021science,clark2015improving}, leading to notable gaps in GHC projections \cite{gardner2009assessing,singh2006effect}. Addressing these challenges involves determining whether significant improvements can be achieved through comprehensive incorporation of model processes, critical parameterization, and finer resolution, but the combined impact of these three advancements in watershed hydrology has not been thoroughly analyzed in large watersheds. Here, we show that a targeted advance in each of these categories can lead to improvements that align with a mechanistic understanding of hydroclimatology. Our analysis of the historic performance of ESMs over the last two generations reveals that mean runoff projections in the current generation have statistically significantly improved over the previous generation, with 95\% confidence bounds, particularly in the larger watersheds of the world. Currently, countries with low GDP per capita and low Human Development Index (HDI) are being impacted by decreasing runoff with heavy population, and future projections also indicate that 40\% of the world's largest river basins could experience decreasing runoff. Our analysis suggests that further improvements in runoff projections are achievable and may help in the analysis of hydroclimate impacts, inform the design of water resources infrastructures, policies and interventions, and enable risk-informed decisions to mitigate the impacts.

The primary hypothesis of this study is that targeted improvements to ESMs can enhance their ability to generate insights critical for water resources management in the largest river basins around the world. We evaluate runoff simulations  from the latest generation of ESMs, specifically, ensembles members from the Coupled Model Intercomparison Project version 6 (CMIP6) \cite{eyring2016overview}. While prior research has assessed CMIP6 performance for runoff generation and its advancements over the previous generations, specifically CMIP5 \cite{hou2023global, guo2022evaluation, milly2005global, alkama2013detection}, corresponding underlying factors driving these variations remain understudied. Recent literature even suggests a possible degradation in the performance of CMIP6 models in some cases \cite{li2021comparative,guo2022evaluation,wang2022performance}, even raising questions about whether these models are approaching their current limits \cite{maslin2012climate}. Analogies can be drawn from the field of weather forecasting, where periods of relative stagnation in numerical weather prediction modelling were followed by improvements, such as in recent years through the use of artificial intelligence \cite {alley2019advances,bauer2015quiet,charlton2024ai}. The rationale for our hypothesis regarding ESM performance is partially based on these analogies. ESMs simulate physical processes (e.g., atmosphere, ocean, land, sea ice) at varying resolutions and may include optional biophysical and biogeochemical components using diverse parameterizations \cite{doscher2022ec}. 
Previous studies have shown that improved resolution refines dynamical and physical parameterizations across atmospheric, oceanic, land, and sea ice systems, enhancing overall model coupling \cite{dunne2020gfdl,shi2019parameter,dawson2015simulating,adcroft2006methods,held2019structure,zhao2018gfdl}. Additionally, numerous studies from different modelling groups have shown that comprehensive processes and enhanced parameterization are crucial for overall model improvements \cite{bao2020fio,caldwell2019doe,collins2011development,dunne2020gfdl,dufresne2013climate}. Despite the intuitive understanding of the importance of these elements, our study attempts to address a key gap by systematically investigating their roles by examining the physics, parameterizations, and resolutions that the process models have incorporated for runoff projections. We explore which combinations of parameterizations, and resolutions yield the most accurate hydrologic projections, providing valuable insights for future model development and refinement.

Here, we focus on assessing runoff in major river basins, which are crucial due to their relatively massive scales and notable impacts on human populations \cite{mekonnen2016future,hogeboom2020capping}. Our findings show that as of 2020, approximately 2.8 billion people reside within the larger river basins. Moreover, due to the coarse resolution of ESMs, many land surface processes are difficult to resolve in small river basins, so we have selected thirty large  river basins based on average discharge to capture these effects more accurately. We evaluate the performance of both Multi-Model Ensemble (MME) and individual models from CMIP6 and CMIP5, focusing on their statistical alignment with reference runoff datasets and the quantification of associated uncertainties. Runoff projections from CMIP6 and CMIP5 are compared using gridded runoff from Global Runoff Reconstruction (GRUN) \cite{ghiggi2019grun} and ECMWF Reanalysis V5 (ERA5) \cite{munoz2021era5} as benchmarks. Additionally, we explore the potential impacts of future runoff changes under the Shared Socioeconomic Pathways (SSP)-3, "Regional Rivalry" carbon emission scenario and investigate how future population dynamics may be affected by changes in surface runoff. Furthermore, we have highlighted how the discrepancies from the best and worst-performing models' future projections can influence impact assessments.

\subsection*{Watershed Hydrology and Demographics}

To understand the relationship between water availability and population growth, we have studied the hydrological and demographic changes in 30 major river basins across the globe from 1970s to 2010s. We used the Global Runoff Reconstruction (GRUN) \cite{ghiggi2019grun} as gridded ground truth for runoff and the Gridded Population of the World (GPW) data from Columbia University \cite{ciesin2018gpw}  for determining hydrologic and demographic dynamics over the last 40 years. We also focused on the income and HDI data from 2020 to better understand the socioeconomic vulnerabilities in conjunction with changes in runoff. For GDP per capita and HDI, we have used the UNDP Human Development Report \cite{undp2024human} and World Bank data, \cite{worldbank2023gdp} respectively.  Figure \ref{fig:f1} provides a detailed visualization of the differences and trends in runoff alongside changes in population density as well as population count, GDP per capita, and HDI in 2020. Increases in runoff are observed in only three river basins (Orinoco, Paraná, and Zambezi), potentially due to changes in regional precipitation patterns. However, a widespread decline in runoff is evident in 67\% of the world's largest river basins, particularly in tropical regions of Africa and South Asia, where major population increases have been observed.  These regions, characterized by low- to middle-income countries with a low HDI, are especially vulnerable. The primary reasons behind the declining runoff in many major river basins are thought to be climate change and human activities \cite{senbeta2021role,yang2019impacts,guan2021past}. Climate change has led to reduced rainfall in several regions, directly impacting water availability. Additionally, human activities, including changes in land use and land classification, have altered the natural flow of water, further reducing runoff \cite{yang2019impacts,hu2021integrated}. Deforestation, urbanization, and the conversion of natural landscapes for agriculture are important contributors to this issue. This growing population, combined with decreasing runoff, presents a critical challenge for the future, which arguably makes it even more crucial to examine future projections from ESMs to anticipate further changes in runoff and develop adaptive strategies. 

\subsection*{Performance of Earth System Models}

We evaluate the performance of all available CMIP6 models (25) in terms of their historical projections of annual runoff from 1960 to 2005. We compare the MME mean and range of variability across 30 river basins against two reference datasets: Global Runoff Reconstruction and Reanalysis. Additionally, we analyze the MME mean from all available CMIP5 ESMs to provide a comparative framework. We have listed all models used in this study in supplementary table \ref{table:s1} with their modelling centers and grid sizes. Our model selection prioritize diversity, including all available models that provide runoff data, while avoiding redundancy by excluding multiple models from the same modeling institute. We have employed two approaches in our analysis. In the first, we have compared 25 members CMIP6 ensemble with 11 members CMIP5 ensemble. In the second approach, we compared 11 models across both generations of ESMs. While the former comparison is a comprehensive assessment of all relevant information from CMIP6 and CMIP5, the latter is a more focused comparison of how the changes in CMIP6 (compared to CMIP5) impact the credibility of runoff simulations. We have highlighted the results for three major rivers with the highest discharge capacity in figure \ref{fig:f2}
, and supplementary figures \ref{fig:s1} and \ref{fig:s2} present results for both approaches across 30 river basins.

Our findings reveal that the CMIP6 MME struggles to accurately capture interannual variability at the river basin scale, with both reference datasets indicating greater interannual variations than those captured by the models. This discrepancy underscores the challenges of modeling complex hydrological processes at finer scales. However, the CMIP6 models show improvement over CMIP5 in mean runoff for historical projections, with this improvement being statistically significant at the 95\% level when compared against ERA5. This suggests some advancement in model accuracy, particularly in capturing mean annual runoff. When comparing the 25 CMIP6 models with the 11 CMIP5 models, we found that CMIP6 models exhibit a higher spread in their projections for river basins with high discharge, such as the Amazon, Congo, Ganges, Brahmaputra, Orinoco, and Rio De La Plata (supplementary figure \ref{fig:s1}, table \ref{table:s2}). This increased spread suggests that the addition of more models introduce more variability in certain contexts. River basins with lower discharge generally show a smaller spread in CMIP6 model projections, although there are exceptions, such as the Irrawaddy, Mekong, and Yangtze rivers. 

We have quantified uncertainty in runoff projections across river basins by analyzing the variability (an implicit assumption being multi-model variability contributes to uncertainty in projections) within the CMIP5 and CMIP6 model ensembles, as depicted in Figure 3. The violin plots display the distribution of model projections for each river basin, highlighting the median and interquartile ranges. Circles above the plots indicate cases where the model ensembles fail to capture the reference runoff within the projected spread for each basin. When comparing all available CMIP6 models (25) with CMIP5 models (11), we find that CMIP6 models exhibit higher variance in their runoff projections, reflecting greater uncertainty. This increased mean uncertainty in CMIP6 reflect a broader exploration of the model parameter space or a more complex representation of physical processes, potentially offering a wider range of plausible futures. Despite higher uncertainty, CMIP6 models more effectively capture the reference runoff across a greater number of river basins compared to CMIP5 models. On the contrary, when comparing 11 models from each CMIP generation, both ensembles show similar levels of uncertainty; however, the CMIP5 models demonstrate better performance in capturing the reference runoff for more river basins than the CMIP6 models.  {This observation aligns with recent literature \cite{guo2022evaluation, wang2022performance}, which also finds that CMIP6 models, despite their advancements, do not always outperform (in fact, sometimes performs worse than) CMIP5 models in all contexts. One potential explanation is that 8 out of the 11 models in the CMIP6 ensemble incorporate new parameterizations related to cloud physics, which were introduced to improve model projections. However, this  parameterization is known to cause \cite{zelinka2020causes} higher climate sensitivity and greater projected warming, leading the IPCC AR6 to even give these models less weight in warming projections\cite{hausfather2022climate}.  It is plausible to hypothesize that these new parameterizations could also affect hydrological variables, such as precipitation, evaporation, and runoff, potentially making CMIP6 runoff projections less consistent with observed patterns. Further studies are needed to investigate these impacts and refine the models accordingly.}

For model evaluation, we examined six commonly-used metrics: Root Mean Square Error (RMSE), Percent Bias, Nash-Sutcliffe Efficiency (NSE), Kling-Gupta Efficiency (KGE), Conditional Bias, and Pearson's Correlation Coefficient (CC) \cite{methods}. These metrics collectively assess different aspects of model performance, specifically, accuracy (RMSE), bias (Percent Bias and Conditional Bias), model efficiency (NSE and KGE), and the strength of the relationship between model predictions and observed data (CC). These metrics were used to validate the performance of CMIP6 and CMIP5 models against reference datasets, shown in supplementary figure \ref{fig:s3}. The analysis is focused on the 11 models from the same institution in each CMIP generation, ensuring a consistent comparison. The evaluation revealed that CMIP6 models generally outperform CMIP5 models in most river basins, as evidenced by the box plots, which show a better alignment of CMIP6 model projections with the reference data. The results highlight improved accuracy, reduced bias, and enhanced overall performance in CMIP6 models, demonstrating superior skill in replicating historical runoff across diverse hydrological contexts. Overall, CMIP6 MME demonstrates strong performance in several key river basins, including the Amazon, Ganges, Brahmaputra, Orinoco, Rio De La Plata, Paraná, Mississippi, Colorado, Indus, Danube, Yellow, Yukon, and Volga rivers. These findings indicate that while CMIP6 models can increase variability, particularly in larger river basins, they also show promise in improving the accuracy of runoff projections when we carefully select and apply them.

\subsection*{Credibility of Hydrologic Projection}
To assess the credibility of hydrologic projections in CMIP6 models, we developed an aggregated ranking matrix based on six key performance metrics outlined in the previous section. For each river basin, all 25 models were ranked for each metric. These rankings were derived by comparing the performance of each model against two runoff reference datasets. Supplementary figures \ref{fig:s4} and \ref{fig:s5} illustrate the rankings of each model for each river basin in comparison to GRUN and ERA5, respectively, with results visualized using color intensity—darker shades indicating better performance. The rankings revealed that no single CMIP6 model consistently outperforms others across all river basins. However, certain models, such as MRI-ESM2, IPSL-CM6A-LR, CNRM-CM6-1 and CESM2, consistently achieved high ranks, demonstrating robust performance across multiple river systems. Subsequently, we created an aggregated rank by combining the results from all twelve rankings—six metrics compared across two reference datasets \cite{methods}. All metrics were given equal weight in the creation of these aggregated ranks. Supplementary figure \ref{fig:s6} displays these aggregated rankings. This ranking is valuable for both water resource managers and earth system modelers or data analysts. For water resource managers, the rankings provide insight into which models are more reliable for informing mitigation and adaptation strategies in their respective river basins. For modelers, these rankings highlight which models perform better in specific basins, allowing them to focus on targeted improvements.

According to the aggregated ranks, some models outperform others overall. To understand why certain models excel, we analyzed the interplay between the physical science implementations, the incorporation of critical parameterizations for runoff, and the spatial resolution of land surface models within ESMs. We plotted the models from best to worst performance and examined the physical processes they modeled and their corresponding resolutions. Figure \ref{fig:f3} provides a detailed analysis of 25 CMIP6 models, showing that advancements in physical science processes, incorporation of critical runoff parameterizations, and finer spatial resolutions significantly improve runoff projections. A model with comprehensive physical and biogeochemical representation typically includes eight key processes: Atmosphere, Aerosol, Ocean, Land Biogeochemistry, Atmospheric Chemistry, Land Ice, and Sea Ice. We also considered critical parameterizations for runoff, such as  Cloud Feedback Model Intercomparison Project (CFMIP), Land Use Model Intercomparison Project  (LUMIP), Land Surface Snow and Soil Moisture Model Intercomparison Project (LS3MIP), and Global Monsoon Model Intercomparison Project (GMMIP).

Our findings reveal that the top-performing models are distinguished by their incorporation of a greater number of these processes and parameterizations, along with finer spatial resolutions. Panel A of the figure displays the total count of processes and parameterizations considered, while Panel B shows the specific processes and parameterizations incorporated by each model. Notably, almost all models include the four core processes—Land Biogeochemistry, Atmosphere, Ocean, and Sea Ice—so these are not highlighted in Panel B. IPSL-CM6A-LR and CNRM-CM6-1 has performed well despite having 250 km resolution, with the incorporation of critical parameterizations.  Interestingly, CanESM5 performed well despite its coarser 500 km spatial resolution, likely due to its comprehensive inclusion of physical processes. On the other hand, models like BCC-CSM2-MR and CMCC-CM2-SR5, despite having finer resolutions (100 km) and extensive parameterizations, underperformed due to the absence of critical elements such as Aerosol, Atmospheric Chemistry, Ocean Bio-geo Chemistry and Land Ice in their simulations. This analysis highlights the importance of both resolution and comprehensive physical process representation in producing credible hydrologic projections.

\subsection*{Impact on Human Population}

We have examined the potential impact on populations under changes in future runoff, focusing on the projected trends in 30 major river basins globally. The analysis is based on long-term future projections from CMIP6 MME under a certain shared socio-economic pathway scenario, where we estimated the mean and trends for surface runoff from 2017 to 2100. CMIP6 models use SSPs, which are more realistic representations of future world \cite{song2022differences, o2016scenario}. There are 5 SSP scenarios and among them SSP126 and SSP245 denotes a greener world in future whereas SSP 370 and SSP 585 has a higher radiative forcing in future \cite{eyring2016overview}. In this study, SSP 370 situation has been considered as this signifies a forcing level familiar to several unmitigated SSP baselines \cite{keller2018carbon} and this corresponds to a 7 $w/m^2$ radiative forcing \cite{eyring2016overview} during the end of the century.  Furthermore, when we assessed the number of people impacted by changes in runoff across different SSPs within our study area (figure \ref{fig:f4}A), we found that the SSP370 scenario projects the highest population growth.  Panel A of the figure highlights the projected population \cite{wang2021global} in the study areas under five Shared Socioeconomic Pathways (SSPs) for the years 2030, 2050, and 2100. Projections shows nearly 40\% of the total population will be living in the study areas under SSP3 scenario ('Regional Rivalry') highlighting the urgency of understanding the potential impacts of changing hydrological conditions under such a high-risk pathway.

The projections from MME mean from 21 CMIP6 model under SSP 370 indicate that at the end of the century, approximately 47\% of the river basins will experience an increase in their long-term mean runoff, with 60\% of the river basins having increasing trends. Panel B of figure \ref{fig:f4} depicts the runoff mean and trends in these river basins under the SSP 370 scenario, with population density in, 2100 overlaid. Notably, basins such as the Ganges, Brahmaputra, Irrawaddy, and Yukon are expected to show the highest increasing trends, while the Amazon, Orinoco, Colorado, Nelson, Churchill, Volga, Orange, Murray and Danube basins are projected to see a decreasing trend. The majority of densely populated areas are expected to coincide with regions showing an increasing trend in runoff. This represents a different scenario from the past 40 years (1970-2010), during which a widespread decline in runoff has been observed in many river basins, particularly in tropical regions of Africa and South Asia. However, it is important to note that CMIP6 MME models have a tendency to overestimate runoff; historical analysis shows that 22 out of 30 river basins tend to exhibit runoff overestimation. This introduces substantial uncertainty in the projected increasing trends for the majority of river basins, highlighting the need for cautious interpretation and careful planning in the future.

Furthermore, we have assessed the number of people that will be affected by changes in runoff trends. A decreasing trend in runoff could lead to water scarcity, impacting numerous people. According to population projections for 2100 under the SSP3 scenario, approximately 5 billion people will live in the 30 major river basins. Among this, approximately 260 million  people will still be affected by decreasing trends in runoff by 2100. This estimate was derived from a 21-member model ensemble (MME). However, when we consider the 5-member ensembles of the best-performing and worst-performing models (based on historical performance), we observe significant differences in the projected impacts. The best-performing models suggest that 10 out of 30 rivers will experience a decreasing trend, potentially affecting 900 million people. In these cases, there is a high level of model agreement, with 90\% of the river basins showing consistent trends across models. Conversely, the worst-performing models indicate that fewer rivers will face decreasing runoff, impacting 360 million people, with model agreement observed in 76\% of the rivers. This discrepancy highlights the need for careful consideration when interpreting model outputs, as the choice of model can significantly influence projected impacts. Moreover, given the tendency of CMIP6 MME models to overestimate runoff, the actual number of people impacted by decreasing runoff trends could be even higher, underscoring the need for careful consideration in future water resource management and planning.

In conclusion, our comprehensive analysis of CMIP6 and CMIP5 models highlights improvements in hydrologic projections, especially in models that incorporate comprehensive physical processes, critical parameterizations, and finer spatial resolutions. However, models incorporating new cloud physics parameterizations in CMIP6 sometimes performed worse than CMIP5 models, underscoring the need for careful evaluation of model enhancements. We have also demonstrated how future trends in runoff could impact millions of people and how the choice of models can significantly influence the projected changes in runoff as well as the number of affected populations. Furthermore, our analysis primarily focuses on mean annual runoff, rather than extreme events, which are crucial for identifying locations that may experience significantly more or less water, leading to flood or drought situations.  Future work should include a detailed examination of these extremes to better inform water resource management and climate adaptation strategies. Additionally, we observed differences among the two reference runoff datasets used for evaluation and, future research should prioritize accounting for the uncertainties associated with difference in reference datasets to improve the reliability of model evaluations.

Figures:

\begin{figure} % Do NOT use \begin{figure*}
	\centering
	\includegraphics[width=0.6\textwidth]{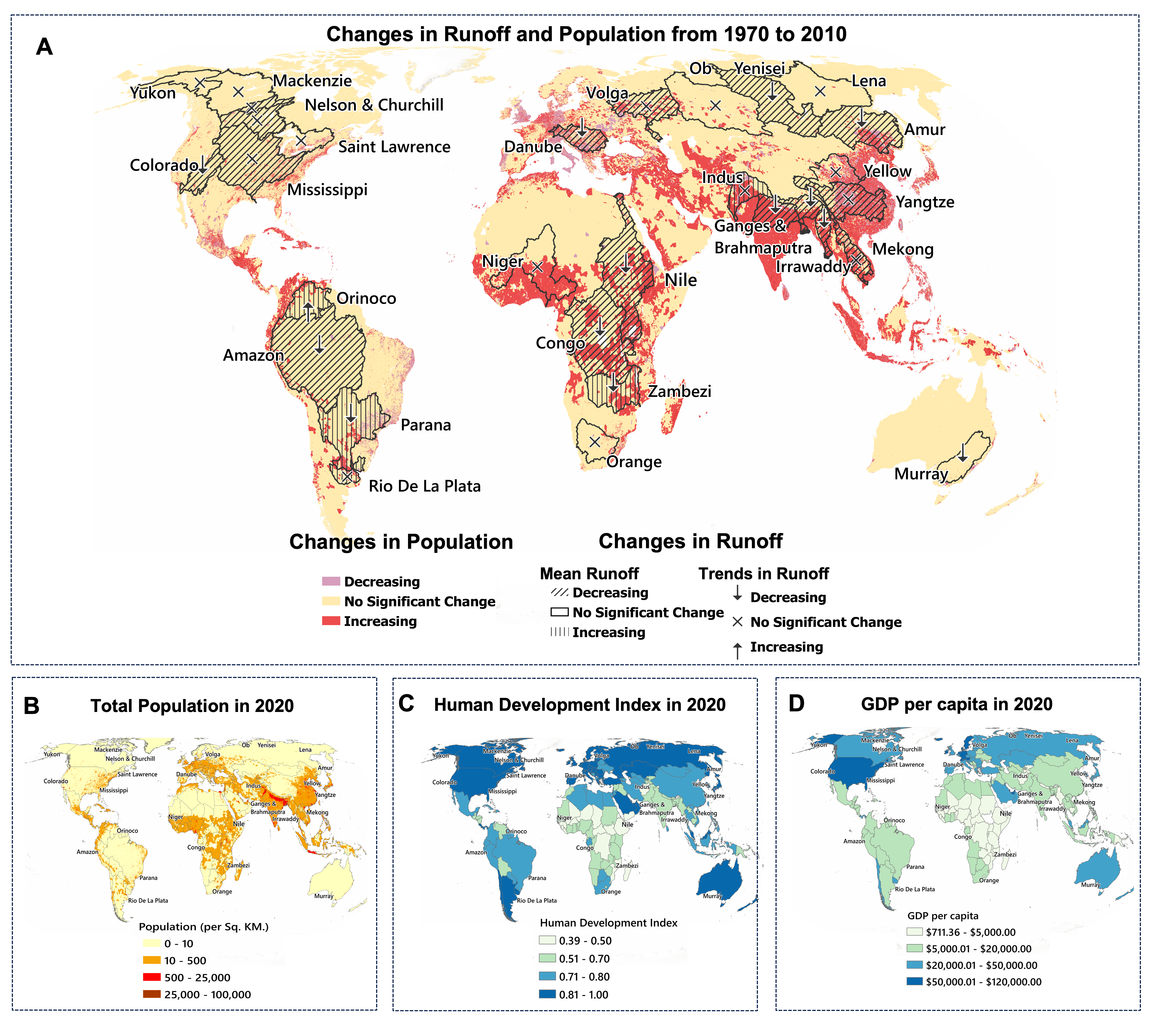} % for an image file named example_figure.*
	% Pick an appropriate width - in print, figures are usually one or two columns wide, which can
	% be approximated by 0.3\textwidth or 0.6\textwidth respectively. Use appropriate label sizes.

	% Captions go below figures
	\caption{\textbf{Decreasing runoff intersects with increasing and more vulnerable population.}
 Analysis of changes in the 2010s (2005-2014) relative to 1970s (1965-1974) in 30 of the largest river basins suggests decreasing runoff intersecting with increasing population, especially in highly populated low- to middle-income countries with a low Human Development Index (HDI).
 (\textbf{A}) Runoff changes are categorized: diagonal hatching for decreases ($< -5$ mm/year), blank for stable ($-5$ to 5$~$mm/year), and vertical hatching for increases ($> 5$ mm/year). Arrows denote trends: downward for decreasing ($< -0.2$ mm/year), cross for stable ($-0.2$ to 0.2 mm/year), and upward for increasing ($> 0.2$ mm/year). Population changes are marked by pink (decreasing), yellow (stable), and red (increasing), indicating shifts of more than 10 people per square kilometer. Notable declines in runoff are noted in tropical regions of Africa and South Asia, with rising population densities. (\textbf{B}) Global population density, (\textbf{C}) Human Development Index, and (\textbf{D}) GDP per capita in 2020 are shown with varying shades.}
	\label{fig:f1} % give each figure a logical label name
\end{figure}
\begin{figure} % Do NOT use \begin{figure*}
	\centering
	\includegraphics[width=0.6\textwidth]{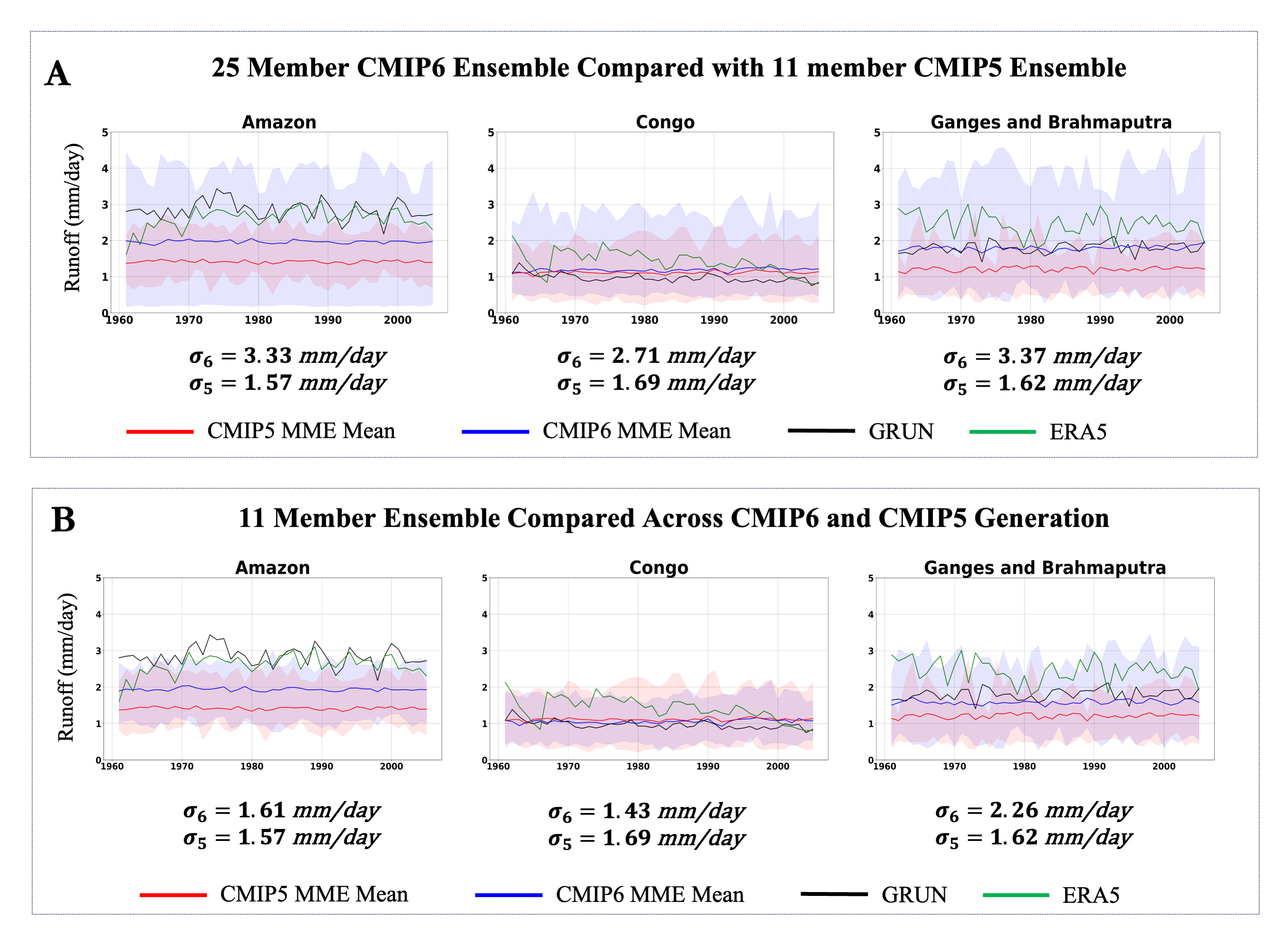} % for an image file named example_figure.*
	% Pick an appropriate width - in print, figures are usually one or two columns wide, which can
	% be approximated by 0.3\textwidth or 0.6\textwidth respectively. Use appropriate label sizes.

	% Captions go below figures
	\caption{\textbf{Improved mean and changed variability for runoff projections in CMIP6 vs CMIP5.}  Three river basins with the highest discharge are shown here, results for 30 basins available in the supplementary section (fig S1-S2). Mean annual runoff and model variability are compared with reconstruction (GRUN) and reanalysis (ERA5) from 1960 to 2005. Red and blue shaded areas represent the spread of CMIP5 and CMIP6 projections, respectively. Solid lines indicate the MME for CMIP5 (red), CMIP6 (blue), GRUN (black), and ERA5 (green). The latest generation of ESMs (CMIP6) shows significant improvement (at 95\%) over CMIP5 in mean runoff for historical projections, when compared against ERA5.  (\textbf{A}) shows Comparative analysis between all available CMIP6 (25) and CMIP5 (11) models without repeating institutions and (\textbf{B}) shows comparison of 11 models which provides runoff projections in both CMIP5 and CMIP6 generation of ESMs. The spread of model projections ($\sigma$) in CMIP6 ESMs seems to be much higher, which can be explained by the addition of higher number of models that generate runoff projections.}
	\label{fig:f2} % give each figure a logical label name
\end{figure}

\begin{figure} % Do NOT use \begin{figure*}
	\centering
	\includegraphics[width=0.6\textwidth]{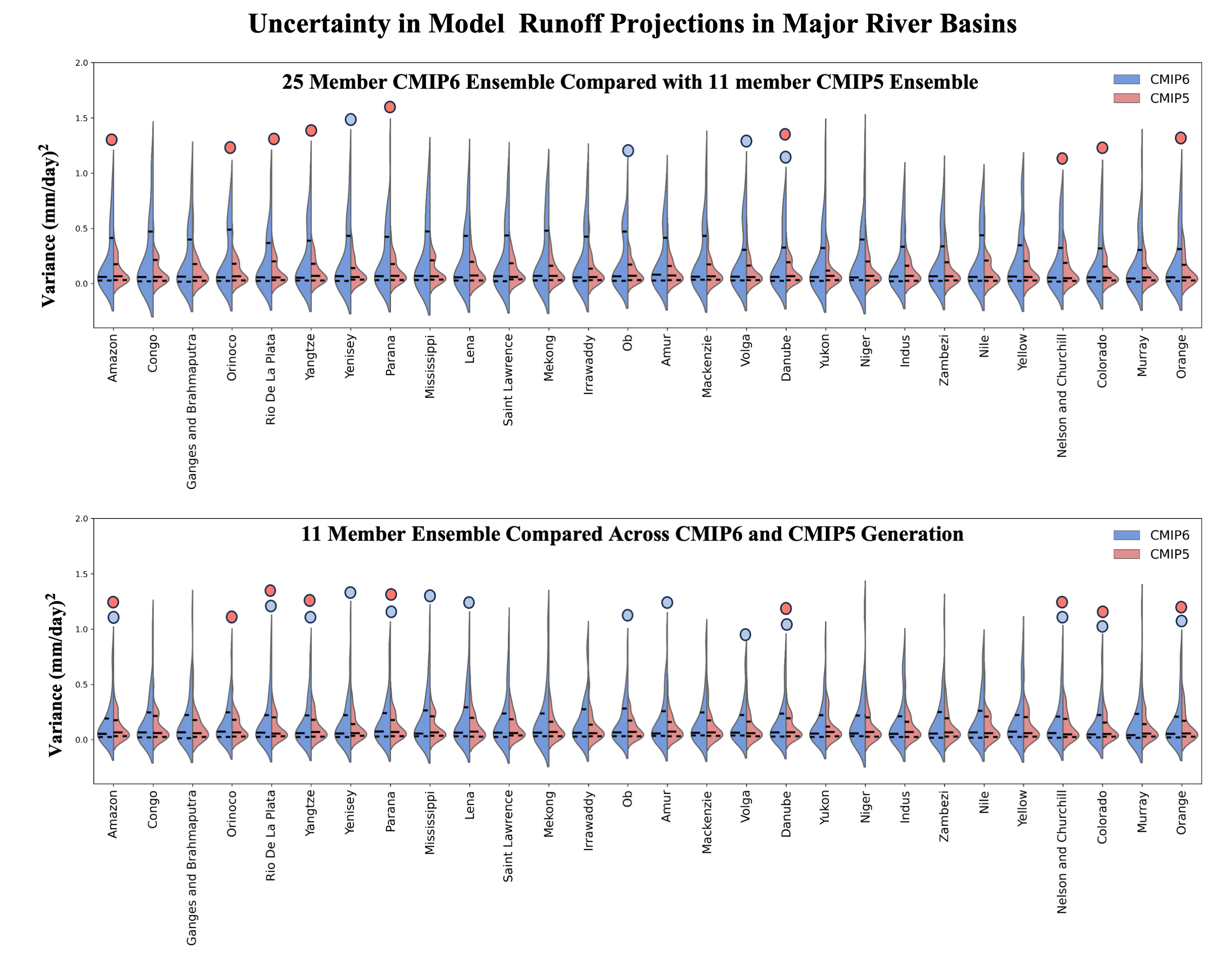} % for an image file named example_figure.*

	% Captions go below figures
	\caption{\textbf{Improved mean runoff projections but larger uncertainty in CMIP6 in comparison to CMIP5.} The violin plots illustrate the variability in model projections for each river basin, with the blue and red shaded areas representing the spread of model projections for CMIP6 and CMIP5, respectively. Solid black lines indicate the median variance, while dashed lines represent the 25th and 75th quartiles. Circles on top of each violin plot indicate whether the ensemble of models failed to capture the reference runoff within the spread of model projections for each river basin. The top panel shows the comparative analysis between all available CMIP6 (25) and CMIP5 (11) models without repeating institutions. The CMIP6 models exhibit a higher variance in their runoff projections across river basins, indicating greater uncertainty in projections and a broader envelope of plausible futures. The bottom panel compares 11 models from each CMIP generation (CMIP5 and CMIP6), demonstrating similar levels of uncertainty in projection but better performance in CMIP5 over CMIP6. This aligns with recent literature \cite{guo2022evaluation,wang2022performance}, but the reasons behind the better performance of CMIP5 models compared to CMIP6 require further investigation.}
	\label{fig:f9} % give each figure a logical label name
\end{figure}

\begin{figure} % Do NOT use \begin{figure*}
	\centering
	\includegraphics[width=0.6\textwidth]{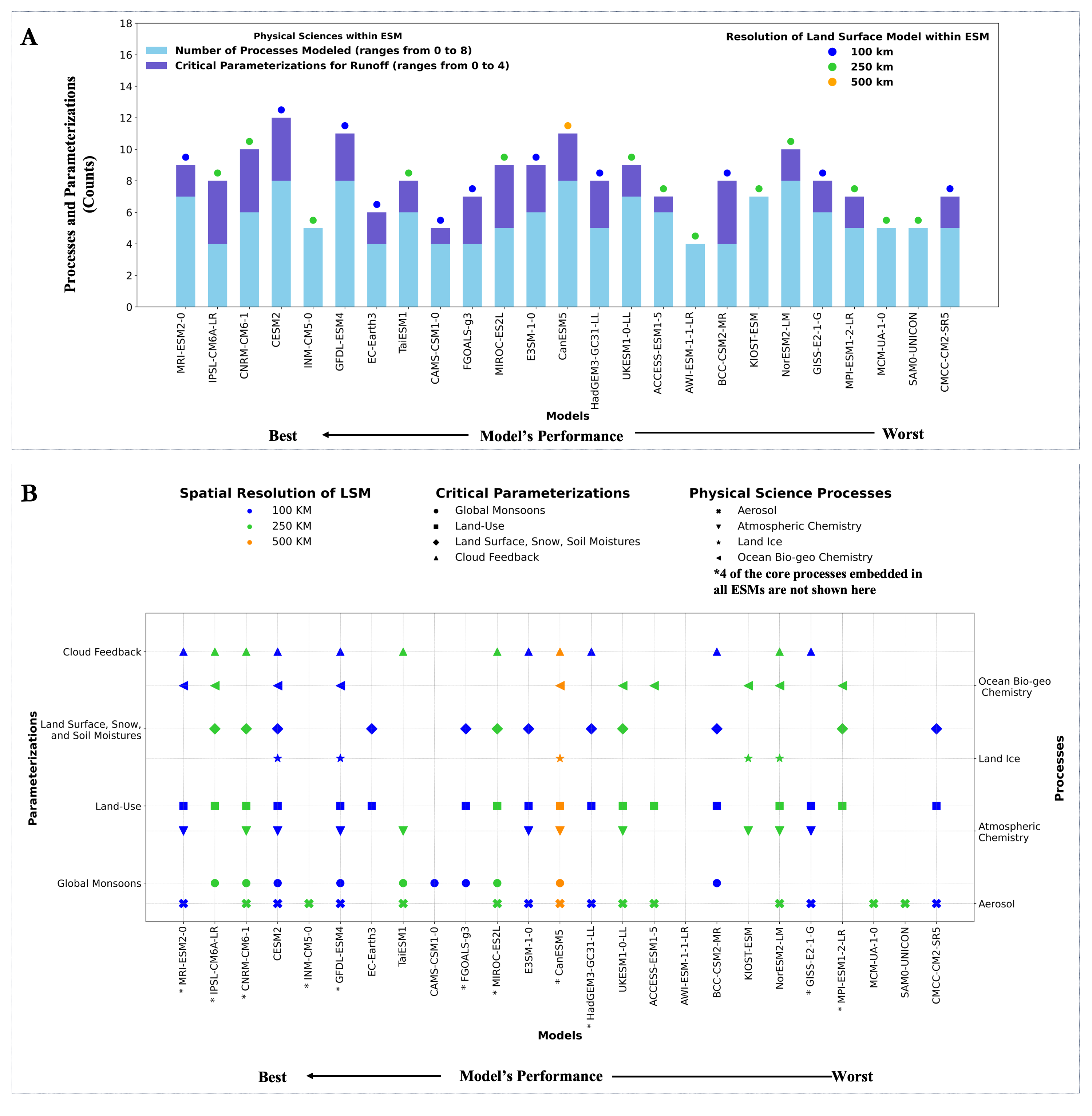} 
	\caption{\textbf{Runoff projections improve with targeted advances in physical science implementation.} 
 An analysis of 25 CMIP6 models shows that advancements in physical science processes, critical parameterizations for runoff, and finer resolution improve runoff projections. (\textbf{A}) The bar chart shows the count of processes (sky blue) and parameterizations (violet) included in each CMIP6 model. Models are categorized by their spatial resolution, represented by colored dots—blue for 100 km, green for 250 km, and orange for 500 km resolution. The chart highlights that leading models typically incorporate a greater number of processes, parameterizations, and finer spatial resolutions. (\textbf{B}) The scatter plot provides a detailed demonstration of processes, parameterizations, and resolutions. Leading models are characterized by their incorporation of cloud feedback, land use, and land surface snow and soil moisture. A star before a model's name indicates its presence in both the CMIP6 and CMIP5 phases. Models are organized from the highest-performing (left) to the lowest-performing (right).}
	\label{fig:f3} % give each figure a logical label name
\end{figure}

\begin{figure} % Do NOT use \begin{figure*}
	\centering
	\includegraphics[width=0.6\textwidth]{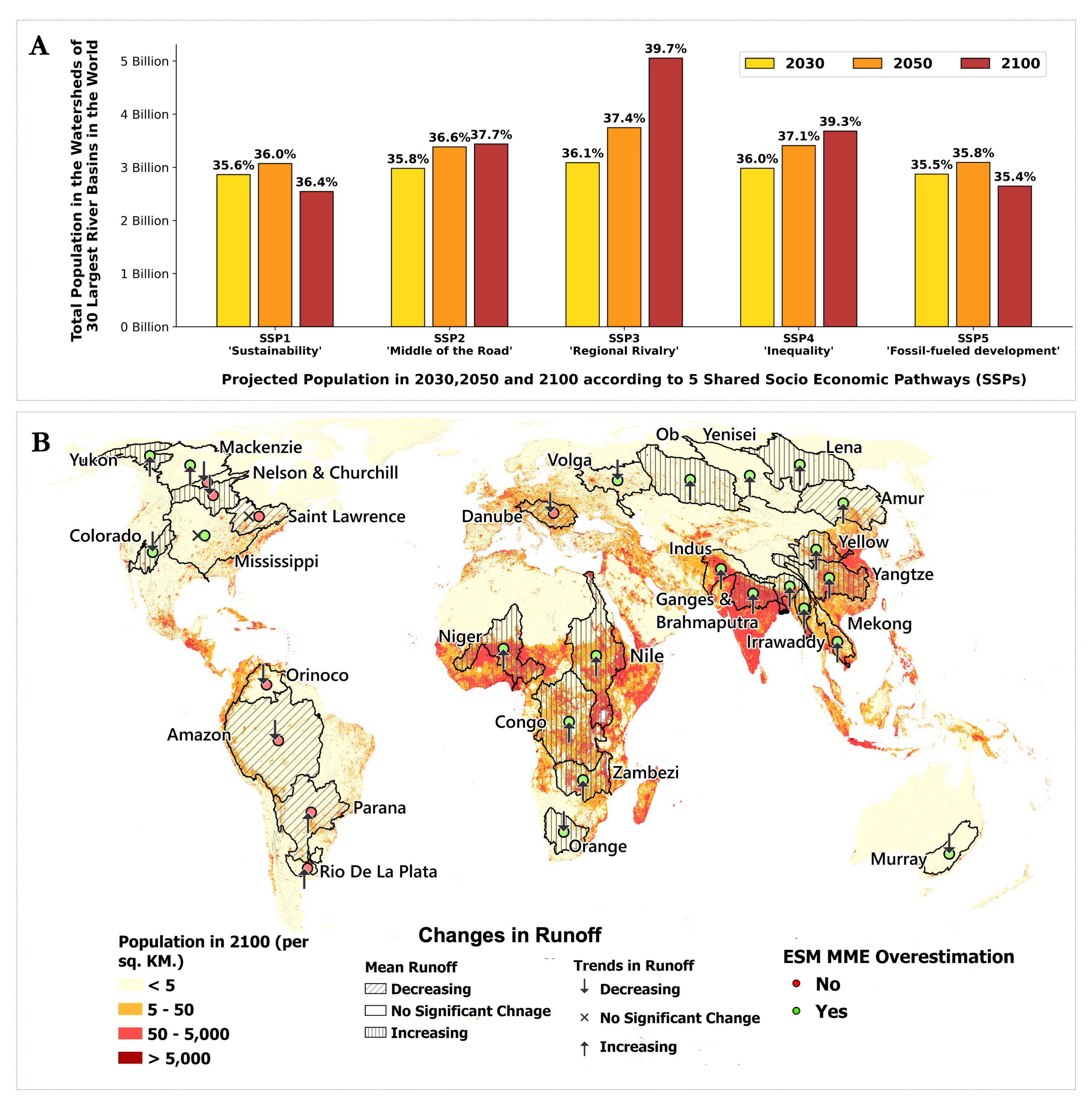} 
	\caption{\textbf{End of century projections suggest that larger population will be at the risk of water deficiency.} 9 of the 30 largest watersheds globally, corresponding to 260 Million people by 2100 according to regional rivalry scenario (SSP3), show decreasing runoff trends. However, historical analysis of CMIP6 MME reveals that 22 of these basins tend to overestimate runoff, indicating even more people could face reduced water availability. (\textbf{A}) Projected population percentages in the study areas under five Shared Socioeconomic Pathways (SSPs) for 2030, 2050, and 2100, with the highest population under SSP3. (\textbf{B}) Changes in mean runoff from historical observations to future projections (CMIP6 MME from 21 Models, SSP 370) and trends in river basins by 2100. Vertical hatches indicate increasing runoff ($> 0.0005$ mm/day); diagonal hatches indicate decreasing runoff ($< 0$ mm/day). Upward arrows denote increasing trends ($> 0.2$ mm/year); downward arrows denote decreasing trends ($< 0$ mm/year). Population density is color-coded, with densely populated areas showing increasing runoff trends. Green and red dots indicate whether CMIP6 MME historically overestimated or underestimated runoff.}
	\label{fig:f4} % give each figure a logical label name
\end{figure}
%%%%%%%%%%%%%%%% MAIN TEXT TABLES %%%%%%%%%%%%%%%

%%%%%%%%%%%%%%%% REFERENCES %%%%%%%%%%%%%%%

\clearpage % Clear all remaining figures and tables then start a new page

% The list of references goes after the main text and before the acknowledgements
% When preparing an initial submission, we recommend you use BibTeX, like this:
%
\bibliography{science_template} % for a file named science_template.bib
\bibliographystyle{sciencemag}

%%%%%%%%%%%%%%%% ACKNOWLEDGEMENTS %%%%%%%%%%%%%%%

\section*{Acknowledgments}
The authors thank current and former members of Northeastern University's Sustainability and Data Sciences Laboratory (SDS Lab) for helpful discussions. 
\paragraph*{Funding:}
The research was supported by DOD SERDP RC20-1183 and partially by NASA Water Resources Program under Grant 21-WATER21-2-0052 (Federal Project ID: 80NSSC22K1138) and the Northeastern University Office of the Provost through AI for Climate and Sustainability (AI4CaS) of the Institute for Experiential AI (EAI). 
\paragraph*{Author contributions:}
PD and ARG conceptualized the research and defined the problem, PD implemented the research and developed the analyses and assessments, PD and ARG interpreted the results and fine-tuned the analyses, assessments, and results, PD prepared the first version of the manuscript while PD and ARG jointly and iteratively developed the final version.
\paragraph*{Competing interests:}
There are no competing interests to declare.
\paragraph*{Data and materials availability:}
CMIP6 and CMIP5 models datasets are available at World Climate Research Program Website hosted by Lawrence Berkeley National Laboratory in the respective addresses (\url{https://aims2.llnl.gov/search/cmip6/}) and (\url{https://aims2.llnl.gov/search/cmip5/}). ERA5 and GRUN runoff data can be obtained from ECMWF Website (\url{https://climate.copernicus.eu/climate-reanalysis}) and here (\url{https://doi.org/10.6084/m9.figshare.9228176}) respectively. Global Runoff Data can be downloaded from (\url{https://www.bafg.de/GRDC/EN/02_srvcs/22_gslrs/gislayers_node.html}). Global Population Data is available at (\url{ https://sedac.ciesin.columbia.edu/}). Future Projection of Global Population Data dataset is available at (\url{https://www.nature.com/articles/s41597-022-01675-x#Sec9}). Human Development Index (HDI) and GDP per capita data are available at (\url{https://ourworldindata.org/grapher/human-development-index} and
(\url{https://ourworldindata.org/grapher/gdp-per-capita-worldbank?time=2022} respectively.

%%%%%%%%%%%%%%%% SUPPLEMENT LIST %%%%%%%%%%%%%%%

% List the contents of your Supplementary Materials, including the numbers of any
% supplementary figures, tables, external data files etc. and any references that are
% cited only in the supplement. In this example, refs. 7-8 are cited only in the supplement.
% Fill out your numbers accordingly and delete any lines that aren't applicable.
\subsection*{Supplementary materials}
Materials and Methods\\
Figs. S1 to S6\\
Tables S1, S2 \\
References \textit{(51-\arabic{enumiv})}\\ % automatically fills out the last reference number
% (filling out the other numbers automatically is possible but fiddly and liable to break)

%%%%%%%%%%%%%%%% END OF MAIN TEXT %%%%%%%%%%%%%%%

\newpage

%%%%%%%%%%%%%%%% START OF SUPPLEMENT %%%%%%%%%%%%%%%

% Figures, tables, equations and pages in the supplement are numbered S1, S2 etc.
\renewcommand{\thefigure}{S\arabic{figure}}
\renewcommand{\thetable}{S\arabic{table}}
\renewcommand{\theequation}{S\arabic{equation}}
\renewcommand{\thepage}{S\arabic{page}}
\setcounter{figure}{0}
\setcounter{table}{0}
\setcounter{equation}{0}
\setcounter{page}{1} % not 0 as \newpage already started a supplementary page
% References continue the numbering from the main text.

%%%%%%%%%%%%%%%% SUPPLEMENT TITLE PAGE %%%%%%%%%%%%%%%

\begin{center}
\section*{Supplementary Materials for\\ \scititle}

% Author list for the supplement
% Indicate the corresponding authors, but do NOT include institutions here
% It would be nice if the template auto-generated this, but doing so is complicated...
Puja Das,
Auroop R. Ganguly$^\ast$,
% we're not in a \author{} environment this time, so use \\ for a new line
\small$^\ast$Corresponding author. Email: a.ganguly@northeastern.com\\

\end{center}

% Fill out the numbers for each type of supplementary material,
% and delete any lines that aren't applicable.
% These are just example numbers that don't match the rest of this template.
\subsubsection*{This PDF file includes:}
Materials and Methods\\
Figures S1 to S6\\
Tables S1, S2 \\

\newpage

%%%%%%%%%%%%%%%% MATERIALS AND METHODS %%%%%%%%%%%%%%%

\subsection*{Materials and Methods}

\subsubsection*{Dataset}
\paragraph{Earth System Model Projections of Runoff}
Monthly Runoff data are collected from CMIP6 and CMIP5 models as well as reference runoff. For the assessment of performance of CMIP6 models, Experiment ID: Historical and SSP 370 (future projection) are selected. All available models with runoff projections ware used in this study, discarding the models with missing data, and multiple models from the same institutions are also not considered. For historical projections, 25 CMIP6 models and 11 CMIP5 models are available, whereas for future projections, 21 CMIP6 models are used in this study.  In Table S1, the list of models used for each case and the name of their modelling group and resolutions are listed. All the CMIP6 models with historical experiment have data from 1850 to 2014 and CMIP5 models have data from 1850 to 2005. For future experiment, data from 2015 to 2100 is available. Models provide runoffs in $kg/m^2s$ unit and they are converted into mm/day unit for ease of calculation. 
\paragraph{Reference Runoff}
For reference surface runoff in historical timescale, this study considered two sources. One is reanalysis dataset and another one is reconstruction based runoff. For reanalysis dataset, runoff datasets are extracted from the European Union's Earth Observation Program (ERA5) \cite{munoz2021era5}. Reanalysis dataset is created from sparsely available observation data combined with data from climate models or remote sensing. Reanalysis datasets are gridded and the grid size (lat x lon) for ERA5 Runoff model is 1800 x 3600.  The dataset was gridded using optimal interpolation. Runoff from ERA5 climate reanalysis dataset was available from 1950 to 2021. Reanalysis datasets are also available at the National Oceanic and Atmospheric Administration (NOAA) \cite{kistler2001ncep}. However, the spatial resolution for NOAA Runoff simulations are 94 x 192, and was very coarse in comparison to the other reanalysis dataset. For these reasons, NOAA dataset was not considered in this study.
For grid-based observations of monthly runoff data, GRUN \cite{ghiggi2019grun} dataset has been used in this study, which is available from 1902 to 2014 with a grid size of 360 x 720. Preprocessing was performed for aligning the coordinates of all models and datasets. To maintain corresponding time frame, 1960-2005 was considered as the historical study period. For future runoff projections, runoff data for the period of 2017-2100 was selected. The spatial information of the rivers was extracted from the Global Runoff Data Centre (GRDC). which can be downloaded from their website (\url{ https://www.bafg.de/GRDC/EN/02_srvcs/22_gslrs/gislayers_node.html}). 
 
\paragraph{Gridded Population Data}

Population data plays a vital role in estimating the impact of changes in runoff in the river basins. Gridded global population (v4) data \cite{ciesin2018gpw} is used in this study for the years 1970, 2010 and 2020. This dataset is available at 1 km spatial resolution and the data are stored in WGS84, geographic coordinate system.  For future population projection in this study, gridded population data was used from a recent study where, population was projected for 5 different SSP scenarios. The population projection was simulated based on the WorldPop dataset and other related covariates using Random Forest algorithm. This dataset is available at 1 km spatial resolution covering 248 countries for a 5-year temporal resolution starting from 2020 to 2100 \cite{wang2021global}. We used population projections for SSP3 scenario in 2020, 2030, 2050, 2070 and 2100. 
For GDP per capita and HDI data we have used, UNDP, Human Development Report (2024) – with minor processing by Our World in Data. and World Bank (2023) – with minor processing by Our World in Data. 

\subsubsection*{Model Performance Metrics}
Model performance and agreement have been assessed using a blend of diagnostic measures and performance metrics, both for a single generation of climate models (from CMIP6) and later for comparisons between different generations (CMIP6 vs CMIP5). This evaluation extends beyond individual model pairs to include assessments of multimodel ensemble (MME) models. The choice of diagnostics and metrics has been guided by the need to evaluate large-scale processes rather than focusing on specific climate phenomena. For this study, some commonly used performance metrics to measure errors, bias and efficiency of projections are used such as Root Mean Squared Error (RMSE), Percent Bias (PBIAS), Conditional Bias (CB), Pearson's Correlation Coefficient (CC), and two hydrologically significant metrics: Nash Sutcliffe Efficiency (NSE) and Modified Kling Gupta Efficiency (KGE'). Model performance is evaluated in comparison to reference runoff dataset (reconstructed observations and reanalysis based). We are specifically comparing mean annual runoff for the period 1960 to 2005, as data from both generations of models is available for this time frame.
 
The RMSE is a very commonly used metric which measures the standard deviation of the residuals or the prediction errors. Again, a low RMSE denote less difference between the observed and simulated variable. RMSE has the same unit as the variable unit, so it provides a better grasp than MSE. 

\begin{equation}
\mathrm{RMSE}= \sqrt{\frac{1}{n}\sum_{i=1}^{n}(x_{i}-y_{i}^2)}
\end{equation}
Where, $n$	=	number of years, $x_{i}$	= reference runoff and $y_{i}$	=	CMIP Projections.

Percent Bias (PBIAS) and Conditional Bias (CB) are widely used metrics for evaluating climate variable estimates. These metrics provide insight into the systematic deviation of model predictions from observed data. PBIAS is particularly informative as it expresses the bias as a percentage, allowing for a clear interpretation of the model’s tendency to overestimate or underestimate the observed values. A low PBIAS is preferred, as it indicates minimal systematic error, while a high PBIAS suggests high model bias, pointing to the model's failure to accurately capture the true values.
\begin{equation}
{\text{PBIAS}} = 100 \times \frac{\sum_{i=1}^{n} \left( x_i - y_i \right)}{\sum_{i=1}^{n} x_i}
\end{equation}

Where, $n$	=	number of years, $x_{i}$	=	reference runoff and $y_{i}$	=	CMIP Projections.

Conditional Bias (CB), on the other hand, offers a more nuanced assessment by examining the bias under specific conditions. It is calculated as the absolute difference between the correlation coefficient of the predicted and observed data, and the ratio of their standard deviations. This approach allows us to understand how well the model’s variability aligns with the observed variability, and how the correlation between the model predictions and observations influences the overall bias. By calculating CB for different observational datasets, we gain insights into the model's performance under varying conditions and its ability to accurately represent the observed climate processes.
\begin{equation}
\text{CB} = \left| r_{\text{Fcst, Obs}} - \frac{S_{\text{Fcst}}}{S_{\text{Obs}}} \right|
\end{equation}
Where, $r_{\text{Fcst, Obs}}$ = correlation coefficient between the projected and reference data, $S_{\text{Fcst}}$ = standard deviation of the projected runoff, $S_{\text{Obs}=}$
standard deviation of the reference runoff.

The Pearson's Correlation Coefficient (CC) is widely used in climate communities for comparison studies. A CC value of 1 denotes the perfect correlation. The formula of CC is as follows:
\begin{equation}
CC =\frac{\sum\left(x_{i}-\bar{x}\right)\left(y_{i}-\bar{y}\right)}{\sqrt{\sum\left(x_{i}-\bar{x}\right)^{2} \sum\left(y_{i}-\bar{y}\right)^{2}}}
\end{equation}
Where,
CC	=	correlation coefficient, $x_{i}$	= reference runoff,  $y_{i}$	=	CMIP Projections, $\bar{x}$	=	mean of the reference runoff values,
$\bar{y}$	=	mean of the CMIP Projections.

Nash Sutcliffe Efficiency (NSE)  is a commonly used and potentially dependable statistic for evaluating predictive skills of hydrologic variables \cite{mccuen2006evaluation}.  NSE value of 1 shows perfect similarity between observed and predicted variable. An NSE value close to 0 means that the mean of the observed values is as good as predicted values. However, negative NSE values mean that the average of the observed data is a better predictor than the simulated data.  The formula is shown below:

\begin{equation}
\mathrm{NSE}= 1- \frac{\sum_{t=1}^{t}((x-y)^2}{\sum_{t=1}^{t}((x-\bar{x})^2}
\end{equation}
Where,
$\mathrm{NSE}$	= Nash Sutcliffe Efficiency, $x$	=	reference runoff at time t, 
$\bar{x}$	=	mean of the reference runoff,
$y$	=	simulated runoff at time t.

NSE value gives an indication of MSE and correlation between observed and predicted variables. But in most cases, measure of variability is also important to understand the efficiency of any variable projections. For this reason, Modified Kling Gupta Efficiency is also estimated in this study as it considered 3 components: correlation coefficient, bias ratio as well as relative variability \cite{kling2012runoff}. Similar to NSE, KGE value of 1 denote perfect simulations. If KGE value is greater than -0.41 then the model simulations are considered to be better than the mean of the observed values.
\begin{equation}
\mathrm{KGE}= 1- \sqrt{(R-1)^2+ (\beta-1)^2 + (\gamma-1)^2}
\end{equation}
\begin{equation}
R =  \frac{Cov (x,y)}{\sigma_x \sigma_y}
\end{equation}
\begin{equation}
\beta  = \frac{\mu_y}{ \mu_x}
\end{equation}
\begin{equation}
\gamma  = \frac{\frac{\sigma_y}{\mu_y}}
{\frac{\sigma_x}{\mu_x}}
\end{equation}
Where,
$\mathrm{KGE}$	= Kling Gupta Efficiency, $R$ = correlation Coefficient, $\beta$ = bias ratio, $\gamma$ = relative variability, $x$	=	observed runoff, $y$	=	simulated runoff,
$Cov (x,y)$ = covariance of reference and simulated runoff, $\sigma_y$	= standard deviation of	simulated runoff, $\mu_y$	=   mean of	simulated runoff, $\sigma_x$	= standard deviation of	observed runoff, $\mu_x$	= mean of	observed runoff.

\subsubsection*{Aggregate Ranking (AR) of Models}

For each river basin, all 25 models are ranked for each of the six metrics from 1 (best performance) to 25 (least performance). For each model $i$, we compute the sum of the ranks across all six metrics:

\begin{equation}
\text{SR}_i = \text{Rank}_{\text{RMSE},i} + \text{Rank}_{\text{PBias},i} + \text{Rank}_{\text{NSE},i} + \text{Rank}_{\text{KGE},i} + \text{Rank}_{\text{CBias},i} + \text{Rank}_{\text{CC},i}
\end{equation}

The Aggregate Ranking (AR) for model $i$ is defined as:
\begin{equation}
\text{AR}_i = \text{SR}_{i}(\text{GRUN}) + \text{SR}_{i}(\text{ERA5})
\end{equation}

where, $\text{SR}_{i}(\text{GRUN})$ is Sum of rankings for model $i$ across all six metrics when compared against the GRUN dataset. $\text{SR}_{i}(\text{ERA5})$ is Sum of rankings for model $i$ across all six metrics when compared against the ERA5 dataset.

To obtain a global or overall score, we average the rankings across all basins for each model:

\begin{equation}
\text{AR} = \frac{1}{n} \sum_{i=1}^{n} \text{AR}_i
\end{equation}

where, $n$: Number of basins considered in the evaluation and $\text{AR}_i$: Aggregate Ranking for model $i$. 
This formulation allows for a comprehensive evaluation of model performance across multiple metrics and datasets. By aggregating the performance scores over different river basins and reference datasets, we obtain a balanced metric for comparing models within CMIP6.

%%%%%%%%%%%%%%%% SUPPLEMENTARY TEXT %%%%%%%%%%%%%%%

% If your supplement is very short you might need to uncomment the following line to avoid
% layout problems with the figures and tables.
%\newpage

%%%%%%%%%%%%%%%% SUPPLEMENTARY FIGURES %%%%%%%%%%%%%%%

\begin{figure} % Do not use \begin{figure*}
	\centering
	\includegraphics[width=0.6\textwidth]{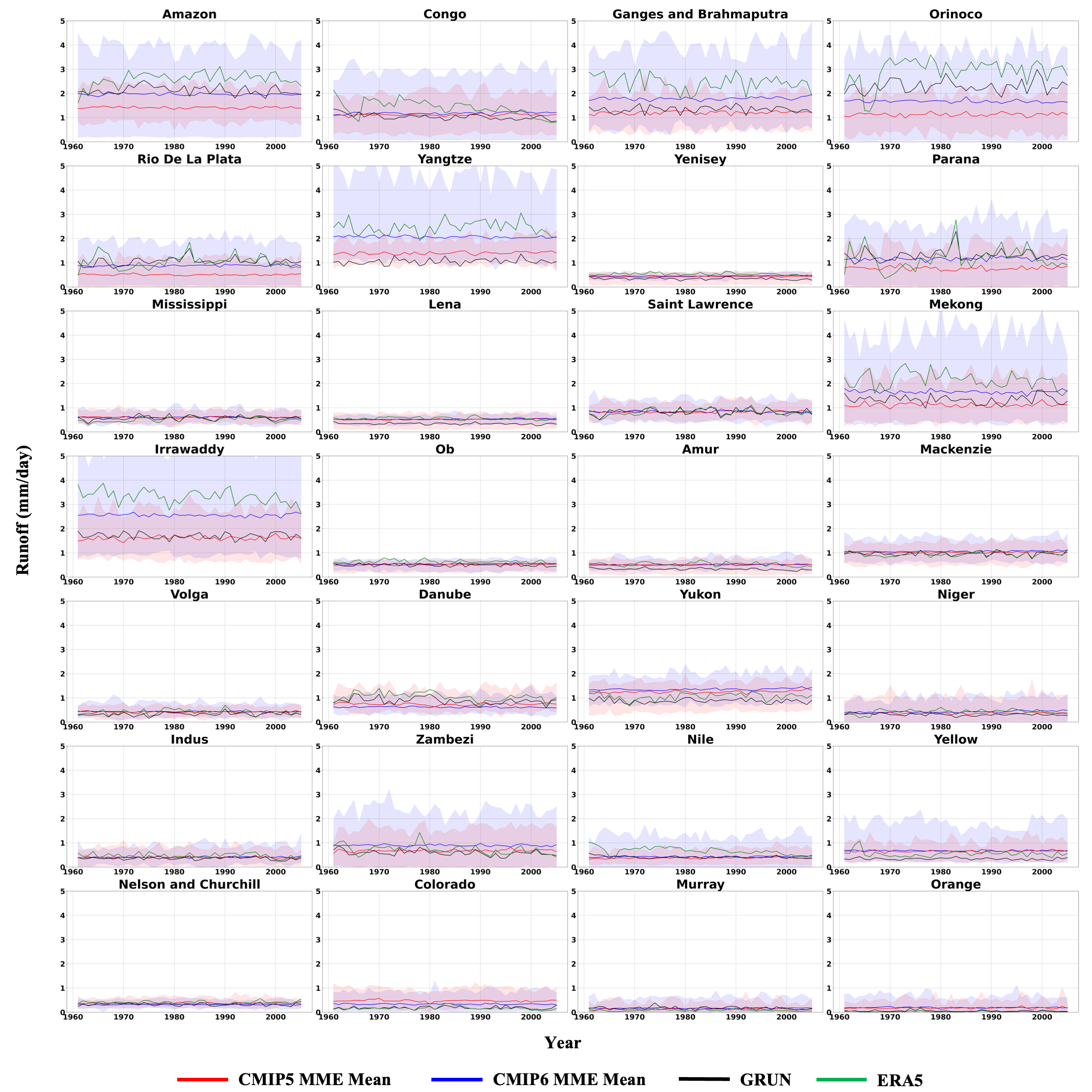} % for an image file named example_figure.*
	% Pick an appriopriate width for the size of the image

	% Captions go below figures
	\caption{\textbf{Mean annual runoff and variability of model projections across generations of ESMs is compared with reconstruction and reanalysis-based Runoffs from 1960 to 2005 in 30 river basins. }
		This is a comparative analysis between all available CMIP6 (25) and CMIP5 (11) models without repeating institutions, The red shaded zones represent the spread of CMIP5 model projections from 11 models, while the blue shaded zones represent CMIP6 projections from 25 models. Solid lines indicate the CMIP5 MME (red), CMIP6 MME (blue), GRUN runoff reconstruction (black), and ERA5 reanalysis (green). The results indicate that variability in larger rivers is generally higher with more CMIP6 models.}
	\label{fig:s1} 
\end{figure}

\begin{figure} % Do not use \begin{figure*}
	\centering
	\includegraphics[width=0.6\textwidth]{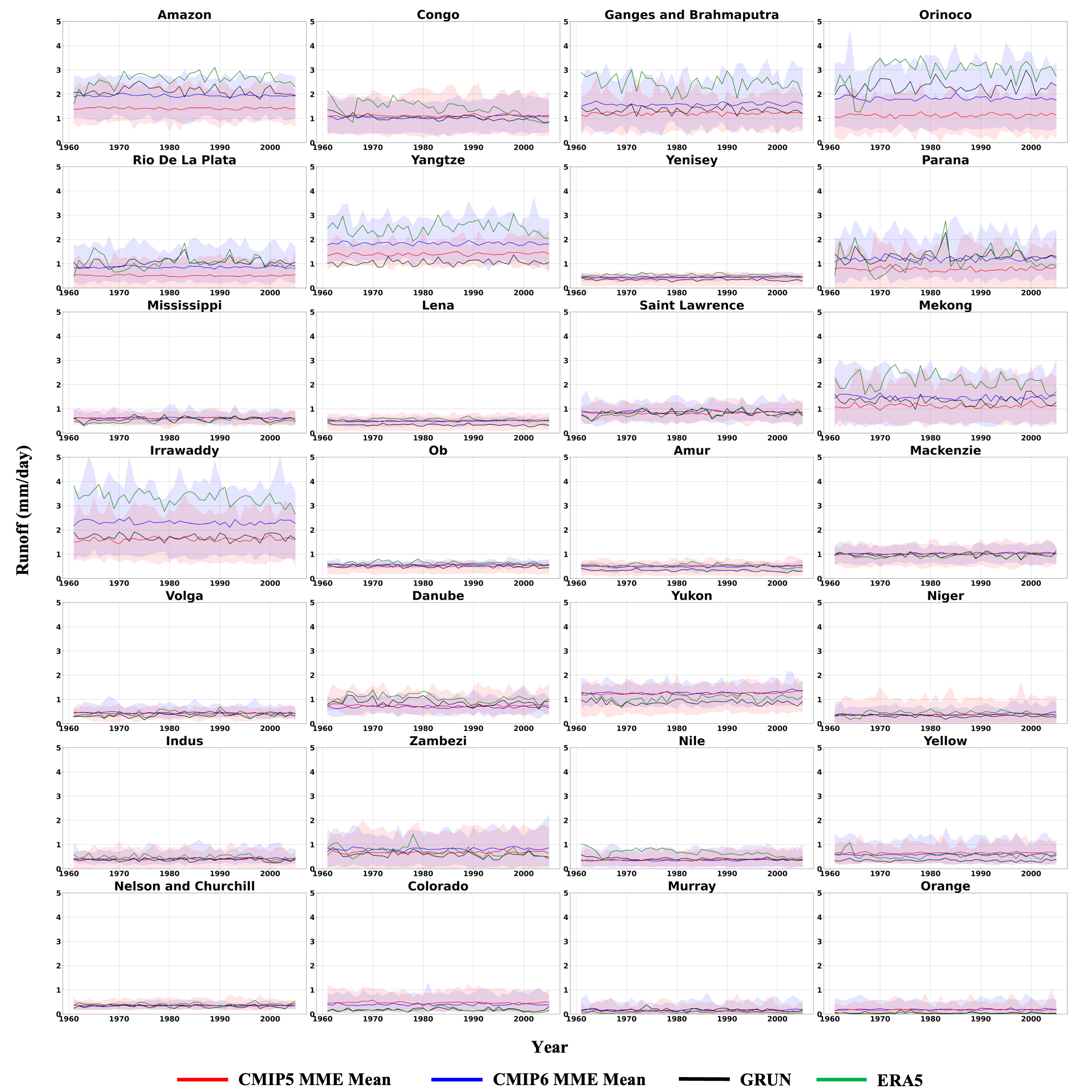} % for an image file named example_figure.*
	% Pick an appriopriate width for the size of the image

	% Captions go below figures
	\caption{\textbf{Mean annual runoff and variability of model projections across generations of ESMs is compared with reconstruction and reanalysis-based runoffs from 1960 to 2005 in 30 river basins. }
		It shows comparison of 11 models across CMIP5 and CMIP6 generation of ESMs. The red shaded zones represent the spread of CMIP5 model projections from 11 models, while the blue shaded zones represent CMIP6 projections from 11 models. Solid lines indicate the CMIP5 MME (red), CMIP6 MME (blue), GRUN runoff reconstruction (black), and ERA5 reanalysis (green).The results indicate that, the variability is often similar or smaller when using the same 11 models across
both CMIP phases.}
	\label{fig:s2} 
\end{figure}

\begin{figure} % Do not use \begin{figure*}
	\centering
	\includegraphics[width=0.6\textwidth]{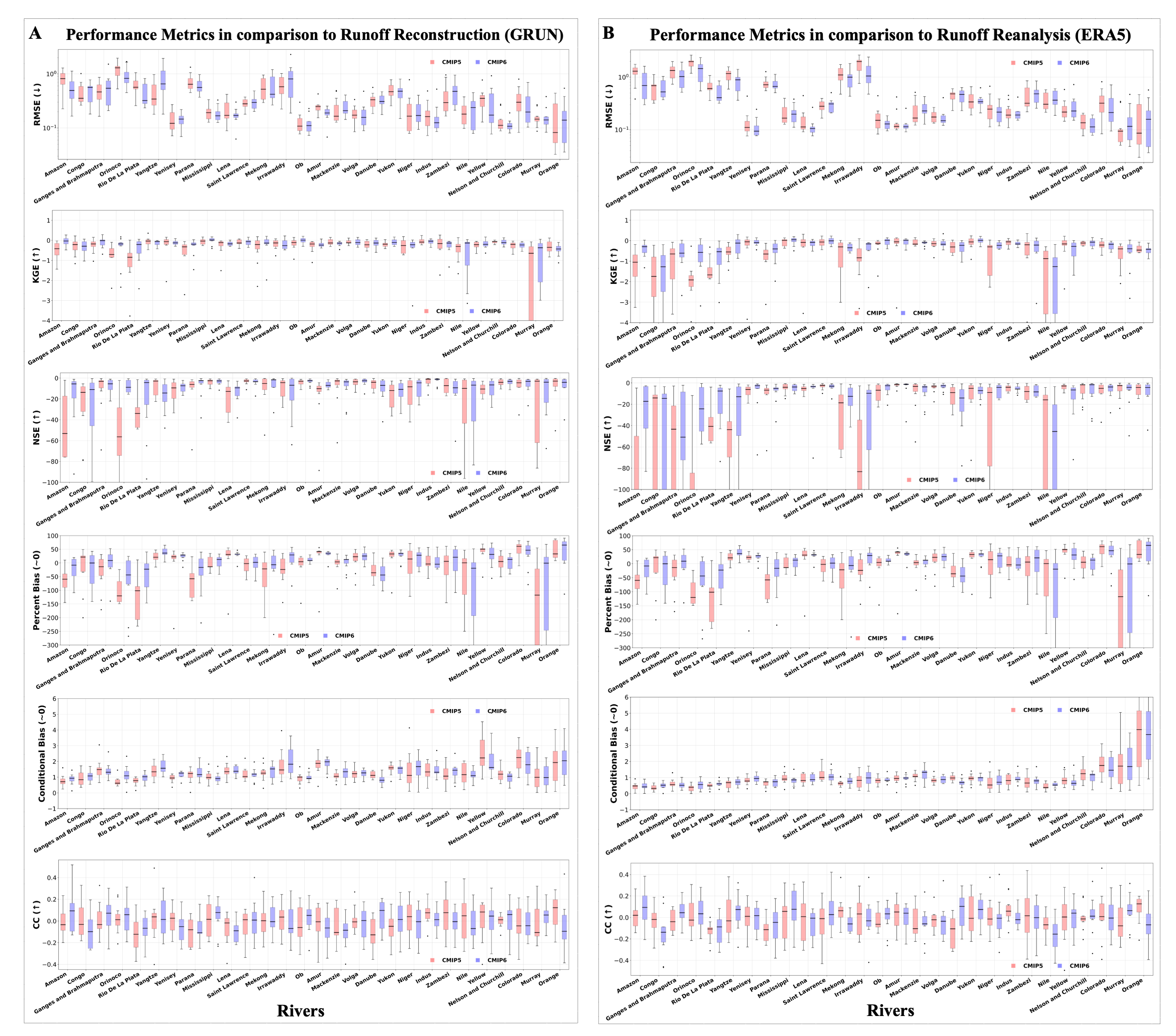} % for an image file named example_figure.*
	% Pick an appriopriate width for the size of the image

	% Captions go below figures
	\caption{\textbf{Performance metrics of CMIP6 and CMIP5 models compared to runoff reconstructions and reanalysis across 30 major rivers in 1961 to 2005. }
		Metrics displayed include RMSE, Percent Bias, NSE, KGE, Conditional Bias, and CC.  $\uparrow$,(\textasciitilde 0)  and $\downarrow$ symbols indicate that, higher score, closer to 0 and lower score is better respectively. In the majority of the river basins, CMIP6, models (shown with blue boxes) performance is better than CMIP5 models (shown with red boxes). (A) shows comparative analysis of performance metrics between CMIP6 and CMIP5 datasets against the Runoff Reconstruction (GRUN) while (B) shows similar comparative analysis, but here the CMIP6 and CMIP5 datasets are compared against the Runoff Reanalysis (ERA5) across the same river basins. Results highlight the variations in accuracy and bias between the model simulations and the reference runoff data.}
	\label{fig:s3} 
\end{figure}
\begin{figure} % Do not use \begin{figure*}
	\centering
	\includegraphics[width=0.6\textwidth]{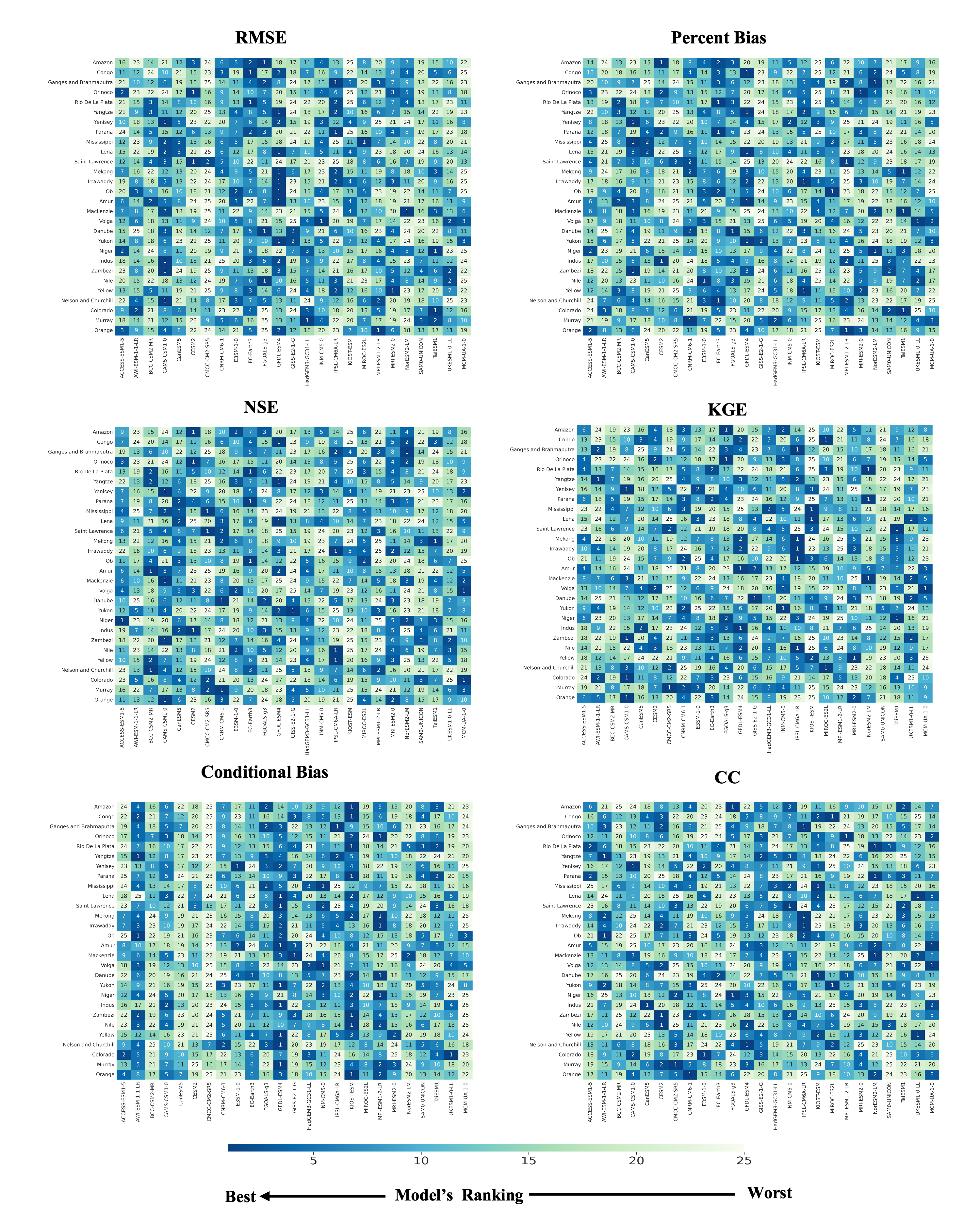} % for an image file named example_figure.*
	% Pick an appriopriate width for the size of the image

	% Captions go below figures
	\caption{\textbf{Ranks for 25 CMIP6 models based on their performance against GRUN in simulating runoff in different river basins according to different skill scores.}
		 Models’ performance is evaluated according to six metrics: Root Mean Squared Error, Pearson's Correlation Coefficient, Mean Squared Error Skill Score, Kling Gupta Efficiency, Percent Bias, and Conditional Bias.  Each cell's color intensity corresponds to the model's rank, with darker shades indicating a higher performance rank. The models are assessed across numerous river basins, with ranks ranging from 1 (best) to 25 (least accurate). }
	\label{fig:s4} 
\end{figure}
\begin{figure} % Do not use \begin{figure*}
	\centering
	\includegraphics[width=0.6\textwidth]{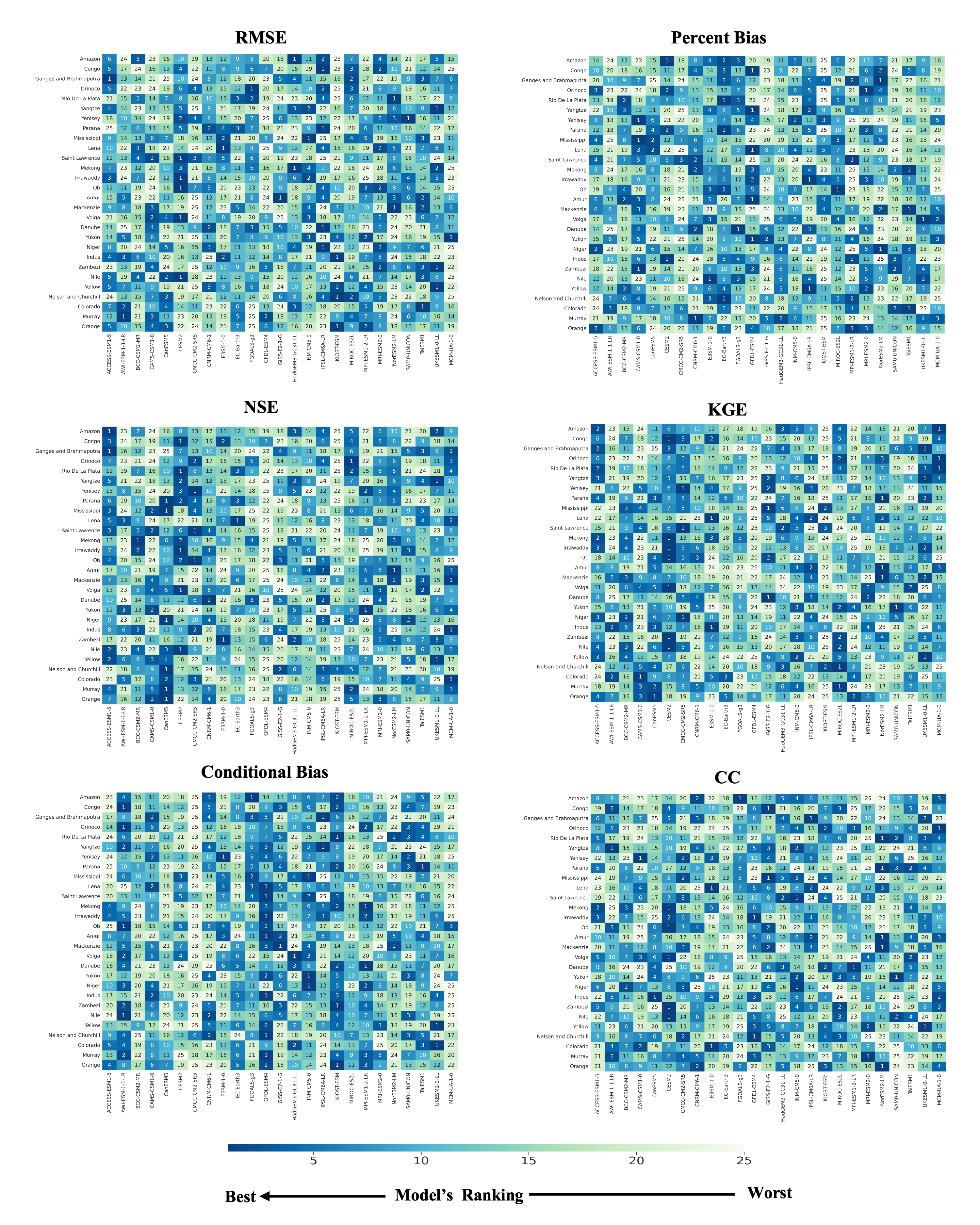} % for an image file named example_figure.*
	% Pick an appriopriate width for the size of the image

	% Captions go below figures
	\caption{\textbf{Ranks for 25 CMIP6 models based on their performance against ERA5 in simulating runoff in different river basins according to different skill scores.}
		Models’ performance is evaluated according to six metrics: Root Mean Squared Error, Pearson's Correlation Coefficient, Mean Squared Error Skill Score, Kling Gupta Efficiency, Percent Bias, and Conditional Bias.  Each cell's color intensity corresponds to the model's rank, with darker shades indicating a higher performance rank. The models are assessed across numerous river basins, with ranks ranging from 1 (best) to 25 (least accurate).}
	\label{fig:s5} 
\end{figure}

\begin{figure} % Do not use \begin{figure*}
	\centering
	\includegraphics[width=0.6\textwidth]{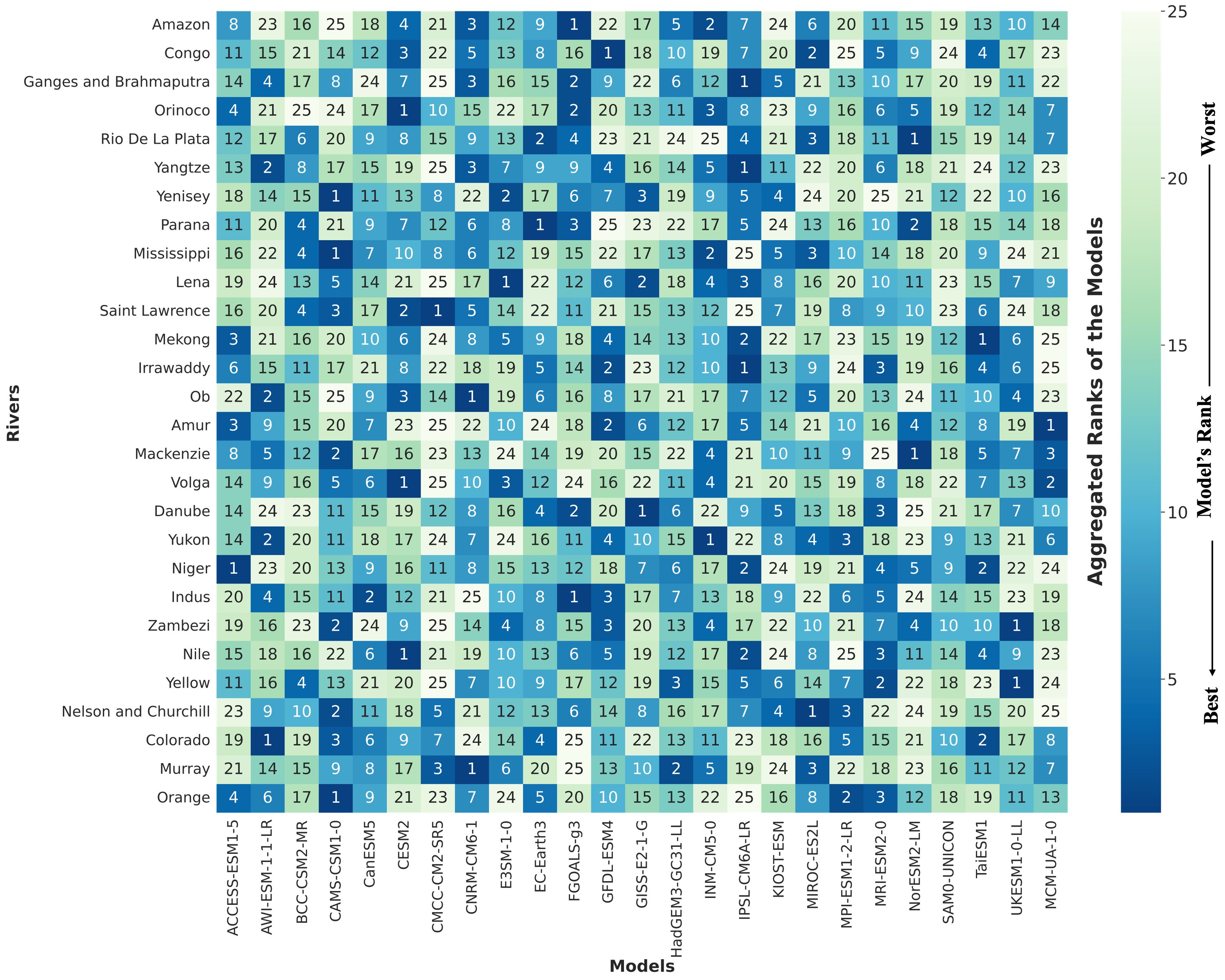} % for an image file named example_figure.*
	% Pick an appriopriate width for the size of the image

	% Captions go below figures
	\caption{\textbf{Aggregated ranks for 25 CMIP6 models based on their performance in simulating runoff in different river basins.}
		 Models’ performance is evaluated according to six metrics: Root Mean Squared Error, Pearson's Correlation Coefficient, Mean Squared Error Skill Score, Kling Gupta Efficiency, Percent Bias, and Conditional Bias. Performance against both reference runoff data is considered while estimating aggregated ranks.  Each cell's color intensity corresponds to the model's rank, with darker shades indicating a higher performance rank. The models are assessed across numerous river basins, with ranks ranging from 1 (best) to 25 (least accurate).}
	\label{fig:s6} 
\end{figure}

%%%%%%%%%%%%%%%% SUPPLEMENTARY TABLES %%%%%%%%%%%%%%%

 % For better looking tables

%\begin{tiny}
%\begin{scriptsize}
\begin{longtable}{|c|p{5.3cm}|p{1.9cm}|p{1.7cm}|p{0.6cm}|p{0.5cm}|p{1.9cm}|p{1.7cm}|}
  \caption{List of IPCC CMIP6 and CMIP5 earth system models with names of modeling centers (or groups), model names, and horizontal grid resolutions used in this study.} 
  \label{tab:commands} \\
  \hline
  \textbf{No} & \textbf{Modeling center} & \textbf{CMIP6 Model Name} & \textbf{CMIP6 Grid Size (lat x lon)} & \textbf{ Hist.} & \textbf{ SSP 370} & \textbf{CMIP5 Model Name} & \textbf{CMIP5 Grid Size (lat x lon)} \\
  \hline
  \endfirsthead
  \caption[]{(Continued)} \\
  \hline
  \textbf{No} & \textbf{Modeling center} & \textbf{CMIP6 Model Name} & \textbf{CMIP6 Grid Size (lat x lon)} & \textbf{ Hist.} & \textbf{SSP 370} & \textbf{CMIP5 Model Name} & \textbf{CMIP5 Grid Size (lat x lon)} \\
  \hline
  \endhead

  \hline
  \endfoot
  \hline
  \endlastfoot 
  1  & Australian Research Council Centre of Excellence for Climate System Science & ACCESS-ESM1-5 & 145 x 192 & \checkmark & \checkmark &  &  \\
  \hline
  2  & Alfred Wegener Institute, Helmholtz Centre for Polar and Marine Research & AWI-ESM-1-1-LR & 96 x 192 & \checkmark & &  &  \\
  \hline
  3  & Beijing Climate Center & BCC-CSM2-MR & 160 x 320 & \checkmark & \checkmark &  &  \\
  \hline
  4  & Chinese Academy of Meteorological Sciences & CAMS-CSM1-0 & 160 x 320 & \checkmark & \checkmark &  & \\
  \hline
  5  & Canadian Centre for Climate Modelling and Analysis & CanESM5 & 64 x 128 & \checkmark & \checkmark & CanESM2 & 64 x 128 \\
  \hline
  6  & Chinese Academy of Sciences & CAS & 128 x 256 &  & \checkmark &  &  \\
  \hline
  7  & National Center for Atmospheric Research & CESM2 & 192 x 288 & \checkmark & \checkmark &  &  \\
  \hline
  8  & Fondazione Centro Euro-Mediterraneo sui Cambiamenti Climatici & CMCC-CM2-SR5 & 192 x 288 & \checkmark & \checkmark &  &  \\
  \hline
  9  & Centre National de Recherches Météorologiques & CNRM-CM6-1 & 128 x 256 & \checkmark & & CNRM-CM5 & 128 x 256 \\
  \hline
  10 & E3SM-Project & E3SM-1-0 & 180 x 360 & \checkmark & &  &  \\
  \hline
  11 & EC-Earth-Consortium & EC-Earth3 & 256 x 512 & \checkmark & \checkmark &  &  \\
  \hline
  12 & Chinese Academy of Sciences & FGOALS-g3 & 80 x 180 & \checkmark & \checkmark & FGOALS-g2 & 60 x 128 \\
  \hline
  13 & National Oceanic and Atmospheric Administration & GFDL-ESM4 & 180 x 288 & \checkmark & \checkmark & GFDL-CM2.1 & 90 x 144 \\
  \hline
  14 & Goddard Institute for Space Studies & GISS-E2-1-G & 90 x 144 & \checkmark & \checkmark & GISS-E2-R & 90 x 144 \\
  \hline
  15 & Met Office Hadley Centre & HadGEM3-GC31-LL & 144 x 192 & \checkmark & & HADCM3 & 72 x 96 \\
  \hline
  16 & Institute for Numerical Mathematics & INM-CM5-0 & 120 x 180 & \checkmark & \checkmark & INM-CM4 & 120 x 180 \\
  \hline
  17 & Institut Pierre Simon Laplace & IPSL-CM6A-LR & 144 x 144 & \checkmark & \checkmark & IPSL-CM5A-LR & 96 x 96 \\
  \hline
  18 & Korea Institute of Ocean Science and Technology & KIOST-ESM & 96 x 192 & \checkmark & &  &  \\
  \hline
  19 & National Institute of Meteorological Sciences & KACE-1-0-G & 144 x 192 & & \checkmark & & \\
  \hline
  20 & Japan Agency for Marine-Earth Science and Technology & MIROC-ES2L & 128 x 256 & \checkmark & \checkmark & MIROC-ESM & 64 x 128 \\
  \hline
  21 & Max Planck Institute for Meteorology & MPI-ESM1-2-LR & 96 x 192 & \checkmark & \checkmark & MPI-ESM-LR & 96 x 192 \\
  \hline
  22 & Meteorological Research Institute & MRI-ESM2-0 & 160 x 320 & \checkmark & \checkmark & MRI-CGCM3 & 160 x 320 \\
  \hline
  23 & NorESM Climate modeling Consortium & NorESM2-LM & 96 x 144 & \checkmark & \checkmark &  & \\
  \hline
  24 & Seoul National University & SAMO-UNICON & 192 x 288 & \checkmark & &  &  \\
  \hline
  25 & Research Center for Environmental Changes & TaiESM1 & 192 x 288 & \checkmark & \checkmark &  &  \\
  \hline
  26 & Met Office Hadley Centre & UKESM1-0-LL & 144 x 192 & \checkmark & \checkmark &  &  \\
  \hline
  27 & Department of Geosciences, University of Arizona & MCM-UA-1-0 & 80 x 96 & \checkmark & \checkmark &  &  
 
  \label{table:s1}
\end{longtable}

\begin{table}[h!]
    \centering
    \caption{Comparison of mean range for each river basin in CMIP6 and CMIP5 MME}
% Adjust the table width to fit within the text width
    \begin{adjustbox}{max width=0.98\textwidth} % Adjust the table width to fit within the text width
    \begin{tabular}{lccc}
        \toprule
        & \multicolumn{3}{c}{\textbf{Mean Range of MME}} \\ % Top row with merged cells
        \cmidrule(lr){2-4} % Line under the top row header
        \textbf{River Basin} & \textbf{ CMIP6 (25 Members)}  & \textbf{CMIP6 (11 Members)}   & \textbf{CMIP5 (11 Members)} \\
        \midrule
        Amazon & 3.334522 & 1.614533 & 1.578802 \\
        Amur & 0.539671 & 0.351061 & 0.627331 \\
        Colorado & 0.849939 & 0.795891 & 0.878067 \\
        Congo & 2.711376 & 1.426638 & 1.690296 \\
        Danube & 0.845414 & 0.791199 & 0.990538 \\
        Ganges and Brahmaputra & 3.368777 & 2.257379 & 1.622380 \\
        Indus & 0.789190 & 0.651129 & 0.790218 \\
        Irrawaddy & 4.055247 & 3.089696 & 2.088546 \\
        Lena & 0.421719 & 0.332031 & 0.659019 \\
        Mackenzie & 0.893737 & 0.690099 & 0.897494 \\
        Mekong & 2.978813 & 2.360596 & 1.877572 \\
        Mississippi & 0.628323 & 0.533670 & 0.588881 \\
        Murray & 0.584074 & 0.496096 & 0.438693 \\
        Nelson and Churchill & 0.373689 & 0.310897 & 0.428368 \\
        Niger & 0.845547 & 0.676674 & 1.107954 \\
        Nile & 1.129224 & 0.777602 & 0.671270 \\
        Ob & 0.365218 & 0.301027 & 0.482278 \\
        Orange & 0.666190 & 0.588270 & 0.522650 \\
        Orinoco & 3.599191 & 2.853824 & 1.675969 \\
        Paraná & 2.587659 & 1.918048 & 1.719214 \\
        Rio De La Plata & 1.791574 & 1.435204 & 1.081464 \\
        Saint Lawrence & 0.956423 & 0.899486 & 0.793997 \\
        Volga & 0.633343 & 0.532918 & 0.533621 \\
        Yangtze & 3.668115 & 1.898487 & 1.146897 \\
        Yellow & 1.616974 & 1.039674 & 0.871159 \\
        Yenisei & 0.308943 & 0.280086 & 0.513663 \\
        Yukon & 1.287718 & 1.005265 & 1.275662 \\
        Zambezi & 2.295283 & 1.454977 & 1.593874 \\
        \bottomrule
      \label{table:s2}
    \end{tabular}
    \end{adjustbox}
\end{table}

\end{document}